\documentclass[10pt,final,doublecolumn]{IEEEtran}
\usepackage{amsmath}
\usepackage{latexsym}
\usepackage{graphicx}
\usepackage{bbding}
\usepackage{enumerate}
\usepackage{multicol}
\usepackage{color}
\usepackage{slashbox}
\IEEEoverridecommandlockouts
\begin{document}

\title{Large-Scale MIMO Relaying Techniques for Physical Layer Security: AF or DF?}
\author{\authorblockN{
Xiaoming~Chen,~\IEEEmembership{Senior Member,~IEEE}, Lei~Lei,~\IEEEmembership{Member,~IEEE,} Huazi~Zhang,~\IEEEmembership{Member,~IEEE,} and Chau~Yuen~\IEEEmembership{Senior Member,~IEEE}
\thanks{Xiaoming~Chen ({\tt chenxiaoming@nuaa.edu.cn}) and
Lei~Lei ({\tt leilei@nuaa.edu.cn}) are with the College of
Electronic and Information Engineering, Nanjing University of
Aeronautics and Astronautics, Nanjing, China. Huazi~Zhang ({\tt
tom.zju@gmail.com}) is with the Department of Information Science
and Electronic Engineering, Zhejiang University, Hangzhou, China.
Chau~Yuen ({\tt yuenchau@sutd.edu.sg}) is with the Singapore
University of Technology and Design, Singapore.}}} \maketitle

\begin{abstract}
In this paper, we consider a large scale multiple input multiple
output (LS-MIMO) relaying system, where an information source sends
the message to its intended destination aided by an LS-MIMO relay,
while a passive eavesdropper tries to intercept the information
forwarded by the relay. The advantage of a large scale antenna array
is exploited to improve spectral efficiency and enhance wireless
security. In particular, the challenging issue incurred by
\emph{short-distance interception} is well addressed. Under very
practical assumptions, i.e., no eavesdropper channel state
information (CSI) and imperfect legitimate CSI at the relay, this
paper gives a thorough secrecy performance analysis and comparison
of two classic relaying techniques, i.e., amplify-and-forward (AF)
and decode-and-forward (DF). Furthermore, asymptotical analysis is
carried out to provide clear insights on the secrecy performance for
such an LS-MIMO relaying system. We show that under large transmit
powers, AF is a better choice than DF from the perspectives of both
secrecy performance and implementation complexity, and prove that
there exits an optimal transmit power at medium regime that
maximizes the secrecy outage capacity.

\end{abstract}


\section{Introduction}
The broadcast nature of the wireless channel gives rise to many
wireless security problems. Traditionally, wireless security is
guaranteed through high layer encryption. With the development of
interception techniques, encryption becomes more complex, leading to
high computation burden. Thanks to physical layer security measures
enlightened by information theory, we may exploit wireless channel
characteristics, i.e., fading and noise, to guarantee secure
communications \cite{Wyner}-\cite{PLS2} even without encryption.
Thus, the heavy dependence on complex higher-layer encryption may be
dispensed, leaving more computation resources for communications.

\subsection{Related Works}
The performance of physical layer security is usually measured by
secrecy rate, namely the capacity difference between the legitimate
channel (from the information source to the destination) and the
eavesdropper channel (from the information source to the
eavesdropper) \cite{SC1}-\cite{SC3}. The introduction of relay into
physical layer security can improve the legitimate channel capacity
through cooperative diversity, and thus enhances transmission
security \cite{Relay1}-\cite{Relay3}. Some feasible relaying schemes
and their performances were discussed in \cite{RelayingSchemes}.
Similar to the traditional relaying systems, amplify-and-forward
(AF) \cite{AF} and decode-and-forward (DF) \cite{DF} are two
commonly used relaying schemes in secure relaying systems.

Intuitively, the position of relay has a great impact on the secrecy
performance. Some optimization schemes for relay position were
presented from the the perspective of minimizing the interception
probability in \cite{Placement}. If there are multiple relays,
higher performance gains can be obtained by relay selection or
cooperation. In \cite{RelaySelection}, an optimal relay selection
scheme was proposed to suppress the interception probability by
exploiting selective gain in multiple relay systems. Additionally,
multiple relay cooperative beamforming combined with jamming was
also adopted to maximize the secrecy rate in \cite{Multi-Relay1} and
\cite{Multi-Relay2}. Considering the CSI sharing overhead, an
iterative distributed precoding was proposed in \cite{Multi-Relay3}
to enhance the received signal power at the destination node and to
mitigate the signal leakage to the eavesdropper simultaneously.

In fact, if the relay is equipped with multiple antennas, it is
likely to enhance the desired signal power and impair the
eavesdropper signal quality simultaneously, and thus increase the
secrecy rate. In this context, multi-antenna relays received
considerable attention, and more effective relaying schemes were
proposed to optimize the secrecy rate \cite{MIMORelay1}
\cite{MIMORelay10}. Linear precoding schemes were investigated in a
MIMO relay network assuming global CSI \cite{MIMORelay2}. Moreover,
two precoding schemes based on the idea of information leakage
neutralization were proposed to enhance wireless security over the
multi-antenna non-regenerative relay assisted multi-carrier
interference channel \cite{MIMORelay3}. However, the quality and
availability of the legitimate and eavesdropper CSI are two research
challenges. Due to the fact that there is a feedback delay or error
for legitimate channel and the eavesdroppers are usually well
hidden, the relay may only have imperfect legitimate CSI and no
eavesdropper CSI. Therefore, it is impossible to realize absolutely
secure communications over the fading channel. To this end, the
conception of secrecy outage capacity was adopted to guarantee
secure communications with a high probability \cite{SOC1}
\cite{SOC2}.

However, without the eavesdropper CSI, a challenging problem rises
when the eavesdropper is close to the transmitter. The so-called
\emph{short-distance interception} scenario is not yet fully
addressed in literature. Even with a multi-antenna relay, the
secrecy outage capacity is too small to fulfill the requirement on
quality of service (QoS), due to the relatively high quality of
interception signal. Recently, it is found that large-scale MIMO
(LS-MIMO) systems can significantly improve the transmission
performance by utilizing its enormous array gain
\cite{LargescaleMIMO1} \cite{LargescaleMIMO2}. Through deploying
hundreds of or even more antennas, the spatial resolution of the
transmit beam can be very high even with just simple linear
precoding methods, e.g., maximum ratio transmission (MRT). Then,
the information leakage to the eavesdropper can be quite small,
and is negligible if the number of antennas approaches infinity \cite{Massive1}-\cite{Massive3}.
In \cite{LS-MIMO}, the achievable secrecy sum-rate based on
regularized channel inversion precoding was derived for
large-scale MISO broadcast channels with confidential message.
Furthermore, large-scale MIMO technology together with artificial
noise (AN) was applied in multicell systems to enhanced wireless
security \cite{LargescaleMIMO3}. It was shown that even with a low
complexity random AN shaping scheme, it was possible to provide a
favorite performance.

Again, the CSI at the LS-MIMO transmitter has a great impact on
the secrecy performance. In general, the CSI in LS-MIMO is
obtained by making use of channel reciprocity of time division
duplex (TDD) systems. Due to duplex delay or pilot contamination,
the estimated CSI may be imperfect, resulting in inevitable
performance loss \cite{PilotContamination}. To the best of our
knowledge, the research on physical layer security based on
LS-MIMO relaying techniques is just beginning, especially in the
case of imperfect CSI. In \cite{ICC14}, we analyzed the secrecy
performance of large-scale MIMO relay in two-hop secure
communications with imperfect CSI, and derived closed-form
expressions of secrecy outage capacity for amplify-and-forward
(AF) and decode-and-forward (DF) relaying protocols, respectively.

\subsection{Main Contributions}
In this paper, we focus on performance analysis and comparison of
two classic relaying schemes, including amplify-and-forward (AF) and
the decode-and-forward (DF), for physical layer security in an
LS-MIMO relaying system with imperfect CSI. The contributions of
this paper are summarized as follows:
\begin{enumerate}
\item We are the first to introduce LS-MIMO techniques into a relaying
system, and solve the challenging problem of short-distance (relay
and eavesdropper) interception.

\item We derive explicit expressions of the secrecy outage capacity
for both AF and DF relaying schemes, and compare their performance
under different conditions, e.g. the CSI accuracy, transmit powers
at the source and the relay, and interception distance.

\item We analyze the asymptotic behavior of secrecy outage capacity
under large transmit power, and obtain some insights on the secrecy
performance as follows:

\begin{enumerate}

\item As the transmit power at the source approaches infinity, the
secrecy outage capacity for both AF and DF relaying schemes is upper
bounded by the same performance ceiling. The bound is is invariant
with respect to transmit powers, but declines as the interception
distance and the outage probability decrease, and increases as the
number of antennas at the relay adds.

\item As the transmit power at the relay approaches infinity, the
secrecy outage capacity of both AF and DF relaying schemes tends to
zero, but the interception event occurs with probability 0 and 1,
respectively.

\item Under large transmit powers, AF is a better choice than DF
from the perspectives of both secrecy performance and implementation
complexity.

\item Both too low and too high relay transmit powers at the relay
result in poor secrecy performance. There exits an optimal transmit
power at medium regime that maximizes the secrecy outage capacity.

\end{enumerate}

\end{enumerate}

\subsection{Paper Organization}
The rest of this paper is organized as follows. We first give an
overview and system model of the LS-MIMO relaying system employing
physical layer security in Section II, and then derive the secrecy
outage capacities for AF and DF relaying schemes with imperfect CSI,
and compare their performance in Section III. In Section IV, we
perform asymptotic analysis to obtain several insights on the
secrecy performance. We then present some simulation results to
validate the effectiveness of the proposed scheme in V. Finally, we
conclude the whole paper in Section VI.

\emph{Notations}: We use bold upper (lower) letters to denote
matrices (column vectors), $(\cdot)^H$ to denote conjugate
transpose, $E[\cdot]$ to denote expectation, $\|\cdot\|$ to denote
the $L_2$ norm of a vector, $|\cdot|$ to denote the absolute value,
and $(a)^{+}$ to denote $\max(a,0)$. The acronym i.i.d. means
``independent and identically distributed", pdf means ``probability
density function" and cdf means ``cumulative distribution function".

\section{System Model}
\begin{figure}\centering
\includegraphics [width=0.45\textwidth] {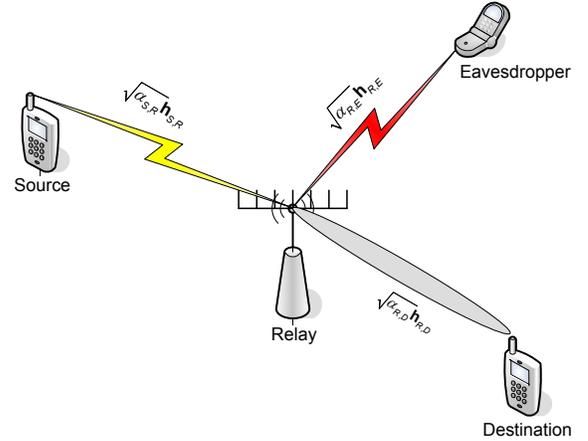}
\caption {An overview of the large-scale MIMO relaying system
employing physical layer security.} \label{Fig1}
\end{figure}

We consider a time division duplex (TDD) LS-MIMO relaying system,
including one source, one destination equipped with a single
antenna each and one relay deploying $N_R$ antennas in presence of
a passive eavesdropper, as shown in Fig.\ref{Fig1}. It is worth
pointing out that $N_R$ is usually quite large in such an LS-MIMO
relaying system, i.e., $N_R=100$ or greater. In addition, the
eavesdropper is usually a legitimate destination or pretends to be
a legitimate destination \cite{Eavesdropper}, so we consider the eavesdropper is
equipped with one antenna, like the destination. The system works
in a half-duplex mode, so the information transmission from the
source to the destination with the aid of the LS-MIMO relay
requires two time slots. Specifically, during the first time slot,
the source transmits the signal to the relay, and then the relay
forwards the post-processing signal to the destination within the
second time slot. Note that the direct link from the source to the
destination is unavailable due to a long distance. Meanwhile, the
eavesdropper monitors the transmission from the relay to the
destination, and tries to intercept the signal. In practice, the
relay is far away from the source, so it is difficult for the
eavesdropper to overhear both the source and the relay. Following
previous related literatures \cite{AF} and \cite{RelaySelection}, we also assume that the eavesdropper only
monitors the relay, since it thought that the signal is from the
relay directly.

We use $\sqrt{\alpha_{i,j}}\textbf{h}_{i,j}$ to denote the channel
from $i$ to $j$, where $i\in\{S,R\}$, $j\in\{R,D,E\}$ with $S, R,
D$ and $E$ representing the source, the relay, the destination and
the eavesdropper, respectively. $\alpha_{i,j}$ is the
distance-dependent path loss and $\textbf{h}_{i,j}$ is the channel
small scale fading. In this paper, we model $\textbf{h}_{i,j}$ as
Gaussian distribution with zero mean and covariance matrix
$\textbf{I}_{N_R}$. $\alpha_{i,j}$ remains constant during a
relatively long period and $\textbf{h}_{i,j}$ fades independently
slot by slot. Thus, the received signal at the relay in the first
time slot can be expressed as
\begin{equation}
\textbf{y}_R=\sqrt{P_S\alpha_{S,R}}\textbf{h}_{S,R}s+\textbf{n}_R,\label{eqn1}
\end{equation}
where $s$ is the normalized Gaussian distributed transmit signal,
$P_S$ is the transmit power at the source, $\textbf{n}_R$ is the
additive Gaussian white noise with zero mean and covariance matrix
$\textbf{I}_{N_R}$ at the relay. Through post-processing to
$\textbf{y}_R$ according to the CSI $\textbf{h}_{S,R}$ and
$\textbf{h}_{R,D}$, the relay forwards a normalized signal
$\textbf{r}$ to the destination with power $P_R$, then the
received signals at the destination and the eavesdropper are given
by
\begin{equation}
y_D=\sqrt{P_R\alpha_{R,D}}\textbf{h}_{R,D}^H\textbf{r}+n_D,\label{eqn3}
\end{equation}
and
\begin{equation}
y_{E}=\sqrt{P_R\alpha_{R,E}}\textbf{h}_{R,E}^H\textbf{r}+n_{E},\label{eqn4}
\end{equation}
respectively, where $n_D$ and $n_{E}$ are the additive Gaussian
white noises with zero mean and unit variance at the destination and
the eavesdropper.

Assuming the achievable legitimate channel rate and the achievable
eavesdropper channel rate are $C_D$ and $C_E$, from the
perspective of information theory, the achievable secrecy rate is
given by $C_{SEC}=[C_D-C_E]^+$, where $[x]^+=\max(x,0)$
\cite{SC1}. Since there is no knowledge of the eavesdropper
channel at the source and relay, it is impossible to maintain a
steady secrecy rate over all realizations of the fading channels.
In this paper, we take the secrecy outage capacity $C_{SOC}$ as
the performance metric, which is defined as the maximum achievable
rate under the condition that the outage probability of the
transmission rate surpasses the secrecy rate is equal to a given
value $\varepsilon$, namely
\begin{equation}
P_r(C_{SOC}>C_D-C_E)=\varepsilon.\label{eqn5}
\end{equation}

\section{Secrecy Performance Analysis and Comparison}
In this section, we concentrate on secrecy performance analysis and
comparison of two commonly used relaying techniques, namely AF and
DF, for physical layer security in an LS-MIMO relaying system with
imperfect CSI. In what follows, we first derive explicit expressions
of the secrecy outage capacity related to a given outage probability
for the two relaying techniques, and then compare their performance
under different channel conditions.

\subsection{Amplify-and-Forward (AF) Case}
In this case, the relay forwards the signal $\textbf{r}^{AF}$ via
multiplying the received signal $\textbf{y}_R$ by a $N_R\times N_R$
processing matrix $\textbf{F}$, namely
\begin{equation}
\textbf{r}^{AF}=\textbf{F}\textbf{y}_R.\label{eqn6}
\end{equation}
We assume the relay has full CSI $\textbf{h}_{S,R}$ through channel
estimation, and gets partial CSI $\textbf{h}_{R,D}$ by making use of
channel reciprocity in TDD systems. Due to duplex delay between
uplink and downlink, there is a certain degree of mismatch between
the estimated CSI $\hat{\textbf{h}}_{R,D}$ and the real CSI
$\textbf{h}_{R,D}$, whose relation can be expressed as
\cite{CSIMismatch}
\begin{equation}
\textbf{h}_{R,D}=\sqrt{\rho}\hat{\textbf{h}}_{R,D}+\sqrt{1-\rho}\textbf{e},\label{eqn7}
\end{equation}
where $\textbf{e}$ is the error noise vector with independent and
identically distributed (i.i.d.) zero mean and unit variance
complex Gaussian entries. $\rho$, scaling from 0 to 1, is the
correlation coefficient between $\hat{\textbf{h}}_{R,D}$ and
$\textbf{h}_{R,D}$. A large $\rho$ means better CSI accuracy. If
$\rho=1$, the relay has full CSI $\textbf{h}_{R,D}$. Additionally,
due to the hidden property of the eavesdropper, the CSI
$\textbf{h}_{R,E}$ is unavailable. Therefore, $\textbf{F}$ is
designed only based on $\textbf{h}_{S,R}$ and
$\hat{\textbf{h}}_{R,D}$, but is independent of
$\textbf{h}_{R,E}$. Since maximum ratio combination (MRC) and
maximum ratio transmission (MRT) can achieve asymptotically
optimal performance in LS-MIMO systems with low complexity
\cite{LargescaleMIMO1} \cite{LargescaleMIMO2}, we design
$\textbf{F}$ by combining MRC and MRT. Specifically, the received
signal $\textbf{y}_R$ is first combined with an MRC vector
$\frac{\textbf{h}_{S,R}^H}{\|\textbf{h}_{S,R}\|}$, then normalized
by a scaling factor
$\frac{1}{\sqrt{P_S\alpha_{S,R}\|\textbf{h}_{S,R}\|^2+1}}$, and
finally multiplied by an MRT vector
$\frac{\hat{\textbf{h}}_{R,D}}{\|\hat{\textbf{h}}_{R,D}\|}$. In
other words, the processing matrix is given by
\begin{equation}
\textbf{F}=\frac{\hat{\textbf{h}}_{R,D}}{\|\hat{\textbf{h}}_{R,D}\|}\frac{1}{\sqrt{P_S\alpha_{S,R}\|\textbf{h}_{S,R}\|^2+1}}\frac{\textbf{h}_{S,R}^H}{\|\textbf{h}_{S,R}\|}.\label{eqn8}
\end{equation}
It is assumed that the destination has perfect CSI by channel
estimation, thus the received signal and the corresponding
signal-to-noise ratio (SNR) can be expressed as
\begin{eqnarray}
y_D^{AF}&=&\frac{\sqrt{P_SP_R\alpha_{S,R}\alpha_{R,D}}\textbf{h}_{R,D}^H\hat{\textbf{h}}_{R,D}\textbf{h}_{S,R}^H\textbf{h}_{S,R}}{\|\hat{\textbf{h}}_{R,D}\|\sqrt{P_S\alpha_{S,R}\|\textbf{h}_{S,R}\|^2+1}\|\textbf{h}_{S,R}\|}s\nonumber\\
&&+\frac{\sqrt{P_R\alpha_{R,D}}\textbf{h}_{R,D}^H\hat{\textbf{h}}_{R,D}\textbf{h}_{S,R}^H}{\|\hat{\textbf{h}}_{R,D}\|\sqrt{P_S\alpha_{S,R}\|\textbf{h}_{S,R}\|^2+1}\|\textbf{h}_{S,R}\|}\textbf{n}_R+n_D,\nonumber\\\label{eqn9}
\end{eqnarray}
and
\begin{eqnarray}
\gamma_D^{AF}&=&\frac{\frac{P_SP_R\alpha_{S,R}\alpha_{R,D}|\textbf{h}_{R,D}^H\hat{\textbf{h}}_{R,D}|^2\|\textbf{h}_{S,R}\|^2}{\|\hat{\textbf{h}}_{R,D}\|^2(P_S\alpha_{S,R}\|\textbf{h}_{S,R}\|^2+1)}}{\frac{P_R\alpha_{R,D}|\textbf{h}_{R,D}^H\hat{\textbf{h}}_{R,D}|^2}{\|\hat{\textbf{h}}_{R,D}\|^2(P_S\alpha_{S,R}\|\textbf{h}_{S,R}\|^2+1)}+1}\nonumber\\
&=&\frac{a|\textbf{h}_{R,D}^H\hat{\textbf{h}}_{R,D}|^2\|\textbf{h}_{S,R}\|^2}{b|\textbf{h}_{R,D}^H\hat{\textbf{h}}_{R,D}|^2+\|\hat{\textbf{h}}_{R,D}\|^2(c\|\textbf{h}_{S,R}\|^2+1)},\label{eqn10}
\end{eqnarray}
where $a=P_SP_R\alpha_{S,R}\alpha_{R,D}$, $b=P_R\alpha_{R,D}$ and
$c=P_S\alpha_{S,R}$. Based on the received SNR in (\ref{eqn10}), we
have the following theorem:

\emph{Theorem 1}: The legitimate channel capacity in an LS-MIMO
relaying system in presence of imperfect CSI can be approximated
as
$C_D^{AF}=W\log_2\left(1+\frac{P_SP_R\alpha_{S,R}\alpha_{R,D}\rho
N_R^2}{P_R\alpha_{R,D}\rho N_R+P_S\alpha_{S,R}N_R+1}\right)$ when
the number of relay antennas is very large, where $W$ is a half of
the spectral bandwidth.

\begin{proof}
Please refer to Appendix I.
\end{proof}

It is found that the legitimate channel capacity is a constant due
to channel hardening in such an LS-MIMO relaying system. Similar
to (\ref{eqn10}), we assume the eavesdropper has perfect CSI by
channel estimation, and thus the SNR is given by
\begin{equation}
\gamma_E^{AF}=\frac{d
|\textbf{h}_{R,E}^H\hat{\textbf{h}}_{R,D}|^2\|\textbf{h}_{S,R}\|^2}
{e|\textbf{h}_{R,E}^H\hat{\textbf{h}}_{R,D}|^2
+\|\hat{\textbf{h}}_{R,D}\|^2(c\|\textbf{h}_{S,R}\|^2+1)},\label{eqn11}
\end{equation}
where $d=P_SP_R\alpha_{S,R}\alpha_{R,E}$ and $e=P_R\alpha_{R,E}$.
Hence, according to the definition of secrecy outage capacity in
(\ref{eqn5}), we have the following theorem:

\emph{Theorem 2}: Given an outage probability bound by
$\varepsilon$, the secrecy outage capacity based on the AF
relaying strategy can be approximated as
$C_{SOC}^{AF}=W\log_2\left(1+\frac{P_SP_R\alpha_{S,R}\alpha_{R,D}\rho
N_R^2}{P_R\alpha_{R,D}\rho
N_R+P_S\alpha_{S,R}N_R+1}\right)-W\log_2\left(1+\frac{P_SP_R\alpha_{S,R}\alpha_{R,E}N_R\ln\varepsilon}{P_R\alpha_{R,E}\ln\varepsilon-P_S\alpha_{S,R}N_R-1}\right)$
when the number of relay antennas is very large.

\begin{proof}
Please refer to Appendix II.
\end{proof}

Based on Theorem 2, we may obtain the interception probability
$P_{0}^{AF}$, namely the probability that the legitimate channel
capacity is less than the eavesdropper channel capacity. By letting
$C_{SOC}^{AF}=0$ in (\ref{app4}), the interception probability can
be computed as (\ref{eqn12}) at the top of the next page.
\begin{figure*}
\begin{eqnarray}
P_{0}^{AF}&=&1-F\left(2^{C_{D}^{AF}/W}-1\right)\nonumber\\
&=&\exp\left(-\frac{(P_S\alpha_{S,R}N_R+1)\left(\frac{P_SP_R\alpha_{S,R}\alpha_{R,D}\rho
N_R^2}{P_R\alpha_{R,D}\rho
N_R+P_S\alpha_{S,R}N_R+1}\right)}{P_SP_R\alpha_{S,R}\alpha_{R,E}N_R-P_R\alpha_{R,E}\left(\frac{P_SP_R\alpha_{S,R}\alpha_{R,D}\rho
N_R^2}{P_R\alpha_{R,D}\rho
N_R+P_S\alpha_{S,R}N_R+1}\right)}\right).\label{eqn12}
\end{eqnarray}
\end{figure*}

It is found that the interception probability of AF relaying scheme
is independent of the spectral bandwidth $W$.

\subsection{Decode-and-Forward (DF) Case}
Different from the AF relaying scheme, DF decodes the received
signal at the relay, and then forwards the original signal to the
destination, so as to avoid noise amplification. Also by using the
MRC technique at the relay, the channel capacity from the source to
the relay can be expressed as
\begin{eqnarray}
C_{S,R}^{DF}&=&W\log_2(1+P_S\alpha_{S,R}\|\textbf{h}_{S,R}\|^2)\nonumber\\
&\approx&W\log_2(1+P_S\alpha_{S,R}N_R),\label{eqn13}
\end{eqnarray}
where (\ref{eqn13}) holds true because of channel hardening when
$N_R\rightarrow\infty$. Then, the relay performs MRT based on the
estimated CSI $\hat{\textbf{h}}_{R,D}$. The channel capacity from
the relay to the destination is given by
\begin{eqnarray}
C_{R,D}^{DF}&=&W\log_2\left(1+P_R\alpha_{R,D}\left|\textbf{h}_{R,D}^H\frac{\hat{\textbf{h}}_{R,D}}{\|\hat{\textbf{h}}_{R,D}\|}\right|^2\right)\nonumber\\
&\approx&W\log_2\left(1+P_R\alpha_{R,D}\rho
N_R\right),\label{eqn14}
\end{eqnarray}
where (\ref{eqn14}) is obtained similarly to Theorem 1. Thus, the
legitimate channel capacity for the DF relaying scheme can be
expressed as
\begin{eqnarray}
C_{D}^{DF}&=&\min(C_{S,R}^{DF},C_{R,D}^{DF})\nonumber\\
&=&W\log_2\left(1+\min(P_S\alpha_{S,R}N_R,P_R\alpha_{R,D}\rho
N_R)\right).\nonumber\\\label{eqn15}
\end{eqnarray}

It is found that $C_D^{DF}$ is also a constant due to channel
hardening in an LS-MIMO system. Meanwhile, the eavesdropper
intercepts the signal from the relay, the corresponding channel
capacity from the relay to the destination can be computed as
\begin{eqnarray}
C_{R,E}^{DF}=W\log_2\left(1+P_R\alpha_{R,E}\left|\textbf{h}_{R,E}^H\frac{\hat{\textbf{h}}_{R,D}}{\|\hat{\textbf{h}}_{R,D}\|}\right|^2\right).\label{eqn16}
\end{eqnarray}

For the secrecy outage capacity in an LS-MIMO DF relaying system, we
have the following theorem:

\emph{Theorem 3}: Given an outage probability bound by
$\varepsilon$, the secrecy outage capacity for the DF relaying
scheme can be approximated as
$C_{SOC}^{DF}=W\log_2\left(1+\min(P_S\alpha_{S,R}N_R,P_R\alpha_{R,D}\rho
N_R)\right)\\-W\log_2\left(1-P_R\alpha_{R,E}\ln\varepsilon\right)$
when the number of relay antennas is very large.

\begin{proof}
Please refer to Appendix III.
\end{proof}

Similarly, we can also obtain the interception probability in this
case. Letting $C_{SOC}^{DF}=0$ in (\ref{app9}), we have
\begin{eqnarray}
P_0^{DF}\!\!\!\!&=&\!\!\!\!\exp\left(-\frac{2^{C_D^{DF}/W}-1}{P_R\alpha_{R,E}}\right)\nonumber\\
&=&\!\!\!\!\exp\left(-\frac{\min(P_S\alpha_{S,R}N_R,P_R\alpha_{R,D}\rho
N_R)}{P_R\alpha_{R,E}}\right).\label{eqn17}
\end{eqnarray}

It is also found that the interception probability of the DF
relaying scheme is independent of the spectral bandwidth $W$.

\subsection{Performance Comparison between AF and DF}
AF and DF are two commonly used relaying techniques. Their
respective merits have been well studied in literature. However,
less is known about their secrecy performance. In this subsection,
we provide an analytical comparison in terms of secrecy outage
capacity and interception probability, and provide explicit
criterions for choosing between AF and DF.

\subsubsection{Secrecy Outage Probability Criteria}
For a given requirement on the outage probability $\varepsilon$, the
difference between the secrecy outage capacity of AF and that of DF
can be computed as (\ref{eqn18}) at the top of the next page.
\begin{figure*}
\begin{eqnarray}
\Delta C_{SOC}&=&C_{SOC}^{AF}-C_{SOC}^{DF}\nonumber\\
&=&W\log_2\left(\frac{(P_R\alpha_{R,D}\rho
N_R+P_S\alpha_{S,R}N_R+P_SP_R\alpha_{S,R}\alpha_{R,D}\rho
N_R^2+1)(P_R\alpha_{R,E}\ln\varepsilon-P_S\alpha_{S,R}N_R-1)}{(P_R\alpha_{R,D}\rho
N_R+P_S\alpha_{S,R}N_R+1)(P_R\alpha_{R,E}\ln\varepsilon-P_S\alpha_{S,R}N_R+P_SP_R\alpha_{S,R}\alpha_{R,E}N_R\ln\varepsilon-1)}\right)\nonumber\\
&&-W\log_2\left(\frac{1+\min(P_S\alpha_{S,R}N_R,P_R\alpha_{R,D})}{1-P_R\alpha_{R,E}\ln\varepsilon}\right).\label{eqn18}
\end{eqnarray}
\end{figure*}
Then, given a set of channel conditions and transmit powers, it is
easy to choose the better relaying technique according to
(\ref{eqn18}). This result can be explained as follows. AF
forwards the noise from the first hop, which not only degrades the
performance at the destination, but also impair the quality of the
eavesdropper signal. In particular, at low source SNR region, the
noise has a greater impact on the eavesdropper signal than the
legitimate signal. In this case, the noise is helpful to improve
the secrecy performance. However, for DF, the benefit brought be
the first hop noise is eliminated through decoding. In other
words, decoding for the legitimate receiver is also in the
interest of the eavesdropper. Thus, AF may outperform DF in this
case. AF and DF have different secrecy performance under various
conditions. Thus, we propose to switch the two relaying protocols
according to system parameters.

\subsubsection{Interception Probability Criteria}
Interception probability is the outage probability corresponding to
zero secrecy outage capacity. In other words, interception
probability is the lower bound on the outage probability. Then, in
terms of interception probability, the secrecy performance gap is
given by (\ref{eqn19}) at the top of this page.
\begin{figure*}
\begin{eqnarray}
\Delta P_0&=&P_0^{AF}-P_0^{DF}\nonumber\\
&=&\exp\left(-\frac{(P_S\alpha_{S,R}N_R+1)\left(\frac{P_SP_R\alpha_{S,R}\alpha_{R,D}\rho
N_R^2}{P_R\alpha_{R,D}\rho
N_R+P_S\alpha_{S,R}N_R+1}\right)}{P_SP_R\alpha_{S,R}\alpha_{R,E}N_R-P_R\alpha_{R,E}\left(\frac{P_SP_R\alpha_{S,R}\alpha_{R,D}\rho
N_R^2}{P_R\alpha_{R,D}\rho
N_R+P_S\alpha_{S,R}N_R+1}\right)}\right)-\exp\left(-\frac{\min(P_S\alpha_{S,R}N_R,P_R\alpha_{R,D}\rho
N_R)}{P_R\alpha_{R,E}}\right).\nonumber\\\label{eqn19}
\end{eqnarray}
\end{figure*}
Similarly, we can choose the better relaying technique from the
perspective of interception probability according to (\ref{eqn19}).

\section{Asymptotic Analysis}
In the last section, we successfully obtained closed-form analytical
results of the secrecy outage capacity for AF and DF relaying
techniques in an LS-MIMO relaying system with imperfect CSI, and
compared their performance. However, the complex structure of the
expressions prevents us from extracting direct guidelines for system
design. In order to obtain clear insights on the secrecy performance
and hence provide some simple but effective relaying schemes, we
carry out asymptotic performance analysis in two extreme cases,
i.e., large transmit powers $P_S$ and $P_R$. In what follows, we
give a detailed investigation of their asymptotic secrecy
performance.

\subsection{AF Case}

\subsubsection{Large $P_S$}
In the case of high transmit power $P_S$ at the source, we have the
following theorem:

\emph{Theorem 4}: There exits a ceiling effect on the secrecy outage
capacity at high $P_S$. This bound is invariant with respect to
$P_S$ and $P_R$, but declines as the interception distance and the
outage probability decrease, and decreases as the number of antennas
at the relay adds.

\begin{proof}
Please refer to Appendix IV.
\end{proof}

In this case, for the interception probability, we have the
following lemma:

\emph{Lemma 1}: Under high $P_S$, the interception probability is
invariant with respect to $P_S$ and $P_R$, but increases as the
interception distance and the outage probability decrease, and
decreases as the number of antennas adds.

\begin{proof}
From (\ref{app12}), the interception probability for the AF relaying
scheme at high $P_S$ can be obtained by letting $C_{SOC}^{AF}=0$ as
follows
\begin{eqnarray}
P_0^{AF}=\exp\left(-\frac{\alpha_{R,D}\rho
N_R}{\alpha_{R,E}}\right).\label{eqn20}
\end{eqnarray}
It is easy to validate the claims of Lemma 1 from (\ref{eqn20}).
Note that if $N_R$ is large, the interception probability
asymptotically approaches zero.
\end{proof}

\emph{Remarks}: Combining Theorem 4 and Lemma 1, the secrecy
performance, including interception probability and secrecy outage
capacity, can be improved \emph{significantly} by increasing the
number of antennas at the relay $N_R$. Thus, even in the
\emph{short-distance interception} scenario, a high secrecy
performance can be realized by adding antennas, which is a main
advantage of LS-MIMO relaying systems with respect to traditional
relaying systems. Again, this result may be well explained by the
extra-high angular resolution of LS-MIMO which is capable of
creating a strong null-space at an even close-by eavesdropper.

\subsubsection{Large $P_R$}
If the transmit power at the relay $P_R$ is sufficiently large, we
have the following theorem:

\emph{Theorem 5}: At high $P_R$, the secrecy outage capacity is
always equal to zero under arbitrary channel conditions and transmit
powers.

\begin{proof}
Please refer to Appendix V.
\end{proof}

For the interception of the AF relaying scheme, we also have the
following lemma:

\emph{Lemma 2}: Under high $P_R$, the interception probability of
the AF relaying scheme is always equal to zero.

\begin{proof}
Examining (\ref{eqn12}), as $P_R$ tends to infinity, we have
\begin{eqnarray}
P\!\!\!\!\!\!\!\!&&\!\!\!\!\!\!\!\!_{0}^{AF}\nonumber\\
&\approx&\!\!\!\!\!\exp\left(-\frac{(P_S\alpha_{S,R}N_R+1)P_S\alpha_{S,R}N_R}{P_SP_R\alpha_{S,R}\alpha_{R,E}N_R-P_R\alpha_{R,E}P_S\alpha_{S,R}N_R}\right)\nonumber\\\label{eqn22}\\
&=&0\label{eqn23},
\end{eqnarray}
where (\ref{eqn22}) holds true since the denominator of
(\ref{eqn22}) is equal to zero.
\end{proof}

\emph{Remarks}: A high $P_R$ will result in zero secrecy outage
capacity, so one should avoid using very high transmit power at the
relay. In fact, we have the following lemma regarding the transmit
power $P_R$:

\emph{Lemma 3}: The secrecy outage capacity first monotonically
increases and finally degrades as $P_R$ increases from 0 to
infinity. There exists an optimal transmit power $P_R$ at medium
regime.

\begin{proof}
As proved in Lemma 2, the secrecy outage capacity tends to zero at
high $P_R$ regime. Meanwhile, the secrecy outage capacity is zero at
$P_R=0$. Thus, there must an optimal transmit power $P_R$ at medium
regime. Similarly, the interception probability has the same
property.
\end{proof}

\subsection{DF Case}

\subsubsection{Large $P_S$}
In the case of high transmit power $P_S$ at the source, we have the
following theorem:

\emph{Theorem 6}: There exits a bound on the secrecy outage capacity
at high $P_S$. This bound is invariant with respect to $P_S$, but
declines as the interception distance and the outage probability
decrease, and increases as the number of antennas at the relay adds.
If $P_R$ is relatively large, the bound is independent of $P_R$.

\begin{proof}
Please refer to Appendix VI.
\end{proof}

According to Theorem 6, it is easy to derive the interception
probability as follows:
\begin{eqnarray}
P_0^{DF}=\exp\left(-\frac{\alpha_{R,D}\rho
N_R}{\alpha_{R,E}}\right),\label{eqn21}
\end{eqnarray}

\emph{Remarks}: The interception probability decreases as $N_R$
increases. In such an LS-MIMO relaying system, the interception
probability asymptotically approaches zero.

\subsubsection{Large $P_R$}
If the transmit power $P_R$ at the relay is sufficiently large, we
have the following theorem:

\emph{Theorem 7}: At high $P_R$, the secrecy outage capacity is
always equal to zero under arbitrary channel conditions and transmit
powers.

\begin{proof}
Please refer to Appendix VII.
\end{proof}

Moreover, for the interception probability of the DF relaying
scheme, we have the following lemma:

\emph{Lemma 4}: The interception probability of the DF relaying
scheme at high $P_R$ is always equal to 1.

\begin{proof}
Examining (\ref{eqn17}), as $P_R$ tends to infinity, we have
\begin{eqnarray}
P_0^{DF}&=&\exp\left(-\frac{\min(P_S\alpha_{S,R}N_R,P_R\alpha_{R,D}\rho
N_R)}{P_R\alpha_{R,E}}\right)\nonumber\\
&=&\exp\left(-\frac{P_S\alpha_{S,R}N_R}{P_R\alpha_{R,E}}\right)\label{eqn24}\\
&=&1,\label{eqn25}
\end{eqnarray}
where (\ref{eqn25}) holds true since the denominator of
(\ref{eqn24}) tends to infinity.
\end{proof}

Note that there also exists an optimal $P_R$ at medium regime in the
sense of maximizing the secrecy performance, and due to page
limitation, we will study this problem in our future works.

In Tab.\ref{Tab1}, we list the secrecy performance of AF and DF
relaying schemes in an LS-MIMO relaying system, so as to compare
their performance conveniently.

\begin{table}\centering
\caption{Secrecy outage capacity and interception probability for AF
and DF relaying schemes at high transmit power regime.} \label{Tab1}
\scriptsize
\begin{tabular}{|c|c|c|c|}\hline
& & large $P_S$ & large $P_R$  \\
\hline  Secrecy Outage & AF & $W\log_2\left(-\frac{\alpha_{R,D}\rho
N_R}{\alpha_{R,E}\ln\varepsilon}\right)$ & 0 \\
\cline{2-4} Capacity  & DF &
$W\log_2\left(\frac{1+P_R\alpha_{R,D}\rho
N_R}{1-P_R\alpha_{R,E}\ln\varepsilon}\right)$ & 0 \\
\hline
& AF & 0 & 0 \\
\cline{2-4} Interception Probability   & DF & 0 & 1 \\
\hline
\end{tabular}
\end{table}

\subsection{Performance Comparison between AF and DF}
Similarly, we compare the secrecy performance of AF and DF relaying
schemes based on their asymptotic behaviors, so as to provide some
simple and effective guideline for relaying scheme selection in
different scenarios.

\subsubsection{Secrecy Outage Capacity Criteria}
From the perspective of secrecy outage capacity, we have the
following two lemmas regarding the relaying scheme selection:

\emph{Lemma 5}: At high $P_S$ regime and $P_R$ is relatively large,
DF has the same secrecy outage capacity as DF. Since AF has lower
complexity than DF, it is a better choice in this case.

\begin{proof}
It is proved by combining Theorem 4 and 6 directly.
\end{proof}

\emph{Remark}: Due to high $P_S$, the noise at the relay during
the first hop is negligible for AF. Meanwhile if $P_R$ is large
enough, AF can asymptotically achieve the same performance as DF.

\emph{Lemma 6}: Under large $P_R$, the secrecy outage capacity for
both AF and DF tends to zero. So AF is also more appealing.

\begin{proof}
It is easy to prove Lemma 6 by combing Theorem 5 and 7.
\end{proof}

\emph{Remark}: Intuitively, if $P_R$ is sufficiently large, the
capacities of the channels from the relay to the destination and
from the relay to the eavesdropper asymptotically approach the
same values, thus secrecy outage capacity tends to zero.

\subsubsection{Interception Probability Criteria}
Again, we can obtain the following two lemmas regarding the
relaying scheme selection by comparing their interception
probability.

\emph{Lemma 7}: At high $P_S$ regime, the interception probability
for both AF and DF approaches zero, so AF is a better choice.

\begin{proof}
It can be proved by comparing (\ref{eqn20}) and (\ref{eqn21})
directly.
\end{proof}

\emph{Remark}: Due to high $P_S$, the noise at the relay during
the first hop is negligible. Meanwhile, by making use of the large
array gain of LS-MIMO relay, the capacity of legitimate channel is
always greater than that of the eavesdropper channel. Thus, the
interception probability tends to zero.

\emph{Lemma 8}: Under large $P_R$, the interception probability for
AF and DF tends to 0 and 1, so AF is more appealing.

\begin{proof}
It is obtained from (\ref{eqn23}) and (\ref{eqn25}).
\end{proof}

\emph{Remark}: If $P_R$ is sufficiently large, for DF, the
capacity of the channel from the relay to the eavesdropper
asymptotically approaches that of the channel from the relay to
the destination. Thus, the interception probability tends to 1.
However, for AF, the noise at the relay is amplified. By
exploiting the array gain of the LS-MIMO relay, the capacity of
the legitimate channel is always greater than the capacity of the
eavesdropper, so the interception probability tends to zero.

Above all, AF is a better choice at high transmit power regime based
on the criterions of secrecy outage capacity and interception
probability. However, as shown later in simulation results, DF may
be better in the low transmit power region. Hence, there is a
switching point between AF and DF from the perspective of maximizing
the secrecy performance.

\section{Simulation Results}
To examine the accuracy and effectiveness of the theoretical
results for the LS-MIMO AF and DF relaying systems, we present
several simulation results in the following scenarios: we set
$N_R=100$, $W=10$KHz and $\rho=0.9$. The relay is in the middle of
a line between the source and the destination. We normalize the
path loss as $\alpha_{S,R}=\alpha_{R,D}=1$ and use $\alpha_{R,E}$
to denote the relative path loss. For example, if
$\alpha_{R,E}>1$, then the eavesdropper is closer to the relay
than the destination. In addition, we use SNR$_S=10\log_{10}P_S$
and SNR$_R=10\log_{10}P_R$ to represent the transmit
signal-to-noise ratio (SNR) in dB at the source and the relay,
respectively. Note that the theoretical results are drawn
according to the derived theoretical expressions, and the
simulation results are obtained by averaging over 10000 times
Monte Carlo simulations under the same condition.

\begin{figure}\centering
\includegraphics [width=0.45\textwidth] {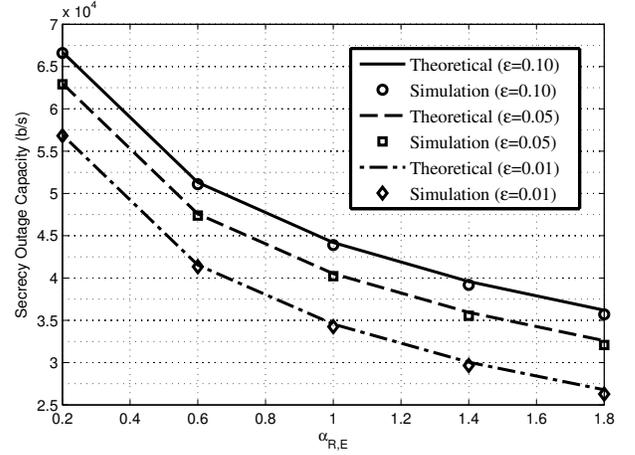}
\caption {Comparison of theoretical and simulation results for the
AF relaying scheme.} \label{Fig2}
\end{figure}

Firstly, we testify the accuracy of the theoretical expression for
the AF relaying scheme with SNR$_S=$ SNR$_R=$ 20dB. As seen in
Fig.\ref{Fig2}, the theoretical results are well consistent with the
simulations in the whole $\alpha_{R,E}$ region with different outage
probability requirements, which proves the high accuracy of the
derived performance expressions. Given an outage probability bound
by $\varepsilon$, as $\alpha_{R,E}$ increases, the secrecy outage
capacity decreases gradually. This is because the interception
capability of the eavesdropper enhances due to the shorter
interception distance. In addition, given an $\alpha_{R,E}$, the
secrecy outage capacity improves with the increase of $\varepsilon$,
since the outage probability is an increasing function of the
secrecy outage capacity.

\begin{figure}\centering
\includegraphics [width=0.5\textwidth] {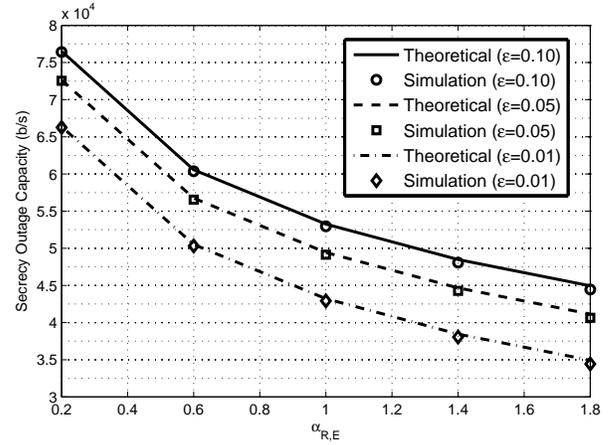}
\caption {Comparison of theoretical and simulation results for the
DF relaying scheme.} \label{Fig3}
\end{figure}

Secondly, we testify the accuracy of the derived theoretical
expressions for the DF relaying scheme with SNR$_s=$ SNR$_R=20$dB.
As seen in Fig.\ref{Fig3}, the theoretical results coincide with the
simulations nicely. Similar to the AF relaying scheme, the secrecy
outage capacity decreases as $\alpha_{R,E}$ increases and
$\varepsilon$ reduces. Note that, compared to the secrecy outage
capacity of the AF relaying scheme in Fig.\ref{Fig2}, the secrecy
outage capacity of the DF relaying scheme is better under the same
conditions, since the DF relaying scheme avoids amplifying the noise
at the relay at low and medium SNR regimes.

\begin{figure}\centering
\includegraphics [width=0.5\textwidth] {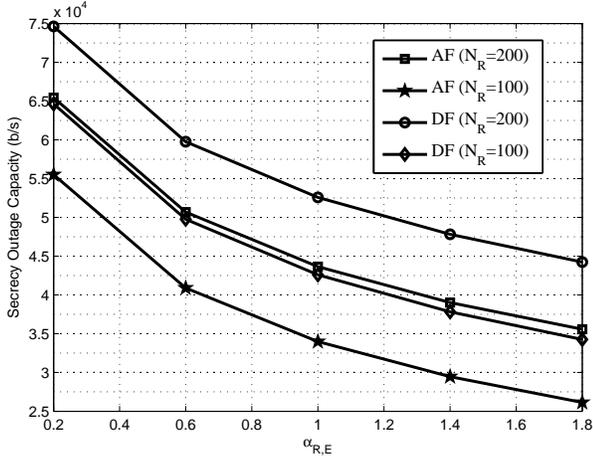}
\caption {Secrecy outage capacity comparison of AF and DF relaying
schemes with different $N_R$.} \label{Fig8}
\end{figure}

Thirdly, we validate the function of $N_R$ on the secrecy outage
capacity of AF and DF relaying schemes with $\varepsilon=0.01$, and
SNR$_S=$ SNR$_R=10$dB. As shown in Fig.\ref{Fig8}, the secrecy
outage capacity can be improved significantly by adding antennas,
even at a large $\alpha_{R,E}$. Thus, LS-MIMO relaying schemes can
effectively solve the problem of short-distance interception. In
addition, it is found that the number of antennas at the relay has
different effects on the secrecy outage capacity for AF and DF
relaying schemes. The secrecy outage capacity of DF with $N_R=100$
is close to that of AF with $N_R=200$ under such condition. Hence,
it makes sense to select the optimal relaying scheme according to
system parameters.

\begin{figure}\centering
\includegraphics [width=0.5\textwidth] {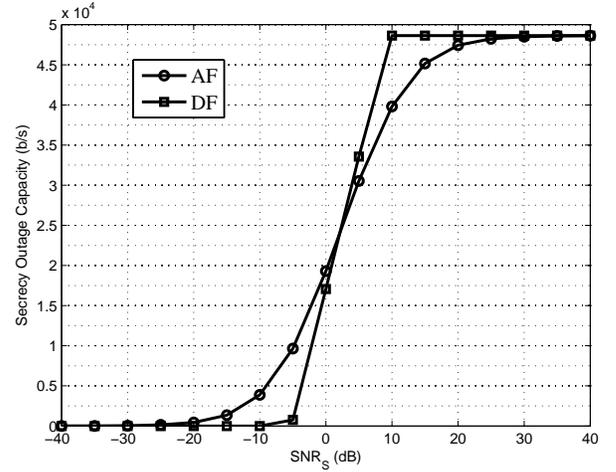}
\caption {Secrecy outage capacity comparison of AF and DF relaying
schemes with different SNR$_S$.} \label{Fig4}
\end{figure}

Fourthly, we show the impact of SNR$_S$ on the secrecy outage
capacity of AF and DF relaying schemes with $\varepsilon=0.05$,
$\alpha_{R,E}=1$, and SNR$_R$=10dB. As seen in Fig.\ref{Fig4}, the
secrecy outage capacity of both AF and DF relaying schemes
increases as SNR$_S$ adds. However, there exits the same
performance bound at high SNR$_S$ regime for both AF and DF, which
confirms the claims in Theorem 4, Theorem 6, and Lemma 5. It is also found
that AF performs better than DF at low SNR$_S$ regime, performs
poorer at medium SNR$_S$ regime, and has the same performance at
high SNR$_S$ regime. This is because at low SNR$_S$, the
forwarding noise of AF may significantly depress the channel
capacity from the relay to the destination, and improve the
secrecy performance. But at medium SNR$_S$, the noise is
negligible, so DF performs better than AF. Therefore, it is better
to switch the two relaying schemes according to system parameters,
so as to maximize the secrecy outage capacity. These cross points
are the switching points indeed.

\begin{figure}\centering
\includegraphics [width=0.5\textwidth] {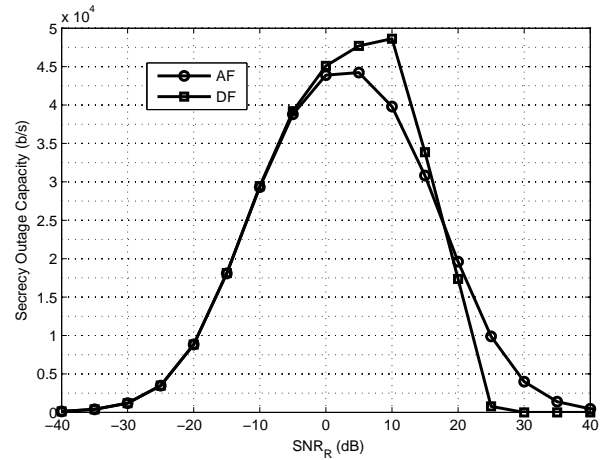}
\caption {Secrecy outage capacity comparison of AF and DF relaying
schemes with different SNR$_R$.} \label{Fig5}
\end{figure}

Then, we investigate the impact of SNR$_R$ on the secrecy outage
capacity of AF and DF relaying schemes with $\varepsilon=0.05$,
$\alpha_{R,E}=1$ and SNR$_S=10$dB. As seen in Fig.\ref{Fig5}, the
secrecy outage capacity is a convex function of SNR$_R$ for both
AF and DF. At low and high SNR$_R$ regimes, the secrecy outage
capacity asymptotically approaches zero, and the maximum secrecy
outage capacity appears at medium SNR$_R$ regime, which proves
Theorem 5, Theorem 7, and Lemma 6 again. Different from the function of
SNR$_S$, AF first performs poorer, since AF would amplify the
noise at the relay. But, AF performs better than DF as SNR$_R$
increases, since the channel capacity from the relay to the
eavesdropper increases and thus the secrecy outage capacity of DF
decreases sharply. Hence, it also makes sense to switch the
relaying schemes according to system parameters.

\begin{figure}\centering
\includegraphics [width=0.5\textwidth] {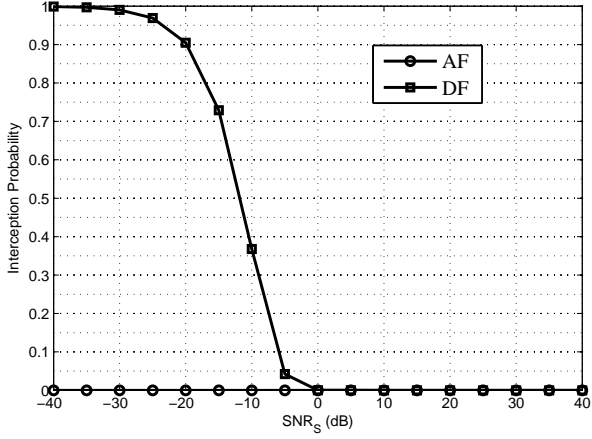}
\caption {Interception probability comparison of AF and DF relaying
schemes with different SNR$_S$.} \label{Fig6}
\end{figure}

Next, we show the impact of SNR$_S$ on the interception probability
of AF and DF relaying schemes with $\alpha_{R,E}=1$, and
SNR$_R=10$dB. It is found from Fig.\ref{Fig6} that AF has a zero
interception probability in the whole SNR$_S$ region, because it can
always support a nonzero rate, as shown in Fig.\ref{Fig3}. At high
SNR$_R$ regime, the interception probability of both AF and DF
approaches a bound independent of $P_S$ and $P_R$, which reconfirms
the claims in Lemma 1 and 7. In this case, due to a large $N_t$, the
bound on the interception probability in (\ref{eqn20}) is nearly
equal to zero.

\begin{figure}\centering
\includegraphics [width=0.5\textwidth] {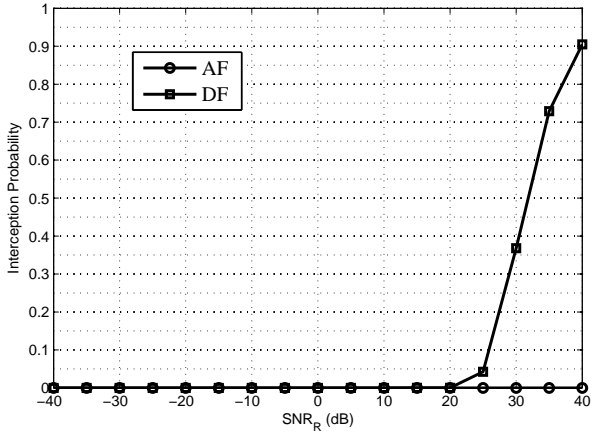}
\caption {Interception probability comparison of AF and DF relaying
schemes with different SNR$_R$.} \label{Fig7}
\end{figure}

Finally, we investigate the impact of SNR$_R$ on the interception
probability of AF and DF relaying schemes with SNR$_S=10$dB and
$\alpha_{R,E}=1$. As seen in Fig.\ref{Fig7}, similar to the last
case, AF has a zero interception probability in the whole SNR$_R$
region. As SNR$_R$ increases, the bounds on the interception
probability for AF and DF are 0 and 1 respectively, which proves
Lemma 2, Lemma 4, and Lemma 8 again.

\section{Conclusion}
A major contribution of this paper is the introduction of LS-MIMO
relaying techniques into physical layer security that leads to
significant enhancement of secrecy outage capacity, even when the
interception distance is short. This paper focuses on the analysis
and comparison of secrecy outage capacity under AF and DF relaying
schemes, and derives explicit expressions in terms of transmit SNRs,
CSI accuracy, antenna number and interception distance. Furthermore,
we carry out asymptotic analysis on the secrecy outage capacity and
interception probability and get clear insights from them.
Altogether, these results provide some important guidelines for the
design of LS-MIMO relaying schemes for secure communications. For
example, AF and DF has the same performance at high source transmit
power. It is better to choose AF due to the lower complexity.
Moreover, there is an optimal relay transmit power at medium regime.

\begin{appendices}
\section{Proof of Theorem 1}
Based on the SNR $\gamma_D^{AF}$ at the destination, the
achievable legitimated channel rate is given by (\ref{app3}) at
the top of the next page,
\begin{figure*}
\begin{eqnarray}
C_D^{AF}&=&W\log_2\left(1+\frac{a|\textbf{h}_{R,D}^H\hat{\textbf{h}}_{R,D}|^2\|\textbf{h}_{S,R}\|^2}
{b|\textbf{h}_{R,D}^H\hat{\textbf{h}}_{R,D}|^2+\|\hat{\textbf{h}}_{R,D}\|^2(c\|\textbf{h}_{S,R}\|^2+1)}\right)\nonumber\\
&=&W\log_2\left(1+\frac{a\left|(\sqrt{\rho}\hat{\textbf{h}}_{R,D}+\sqrt{1-\rho}\textbf{e})^H\frac{\hat{\textbf{h}}_{R,D}}{\|\hat{\textbf{h}}_{R,D}\|}\right|^2
\|\textbf{h}_{S,R}\|^2}{b\left|(\sqrt{\rho}\hat{\textbf{h}}_{R,D}+\sqrt{1-\rho}\textbf{e})^H\frac{\hat{\textbf{h}}_{R,D}}{\|\hat{\textbf{h}}_{R,D}\|}\right|^2
+(c\|\textbf{h}_{S,R}\|^2+1)}\right)\label{app1}\\
&=&W\log_2\left(1+\frac{a\left(\rho\|\hat{\textbf{h}}_{R,D}\|^2+2\sqrt{\rho(1-\rho)}\mathcal{R}(\textbf{e}^H\hat{\textbf{h}}_{R,D})
+(1-\rho)\frac{\|\textbf{e}\hat{\textbf{h}}_{R,D}^H\|^2}{\|\hat{\textbf{h}}_{R,D}\|^2}\right)\|\textbf{h}_{S,R}\|^2}{b\left(\rho\|\hat{\textbf{h}}_{R,D}\|^2
+2\sqrt{\rho(1-\rho)}\mathcal{R}(\textbf{e}^H\hat{\textbf{h}}_{R,D})+(1-\rho)\frac{\|\textbf{e}\hat{\textbf{h}}_{R,D}^H\|^2}{\|\hat{\textbf{h}}_{R,D}\|^2}\right)
+c\|\textbf{h}_{S,R}\|^2+1}\right)\nonumber\\
&\approx&W\log_2\left(1+\frac{a\rho\|\hat{\textbf{h}}_{R,D}\|^2\|\textbf{h}_{S,R}\|^2}{b\rho\|\hat{\textbf{h}}_{R,D}\|^2+c\|\textbf{h}_{S,R}\|^2+1}\right)\label{app2}\\
&\approx&W\log_2\left(1+\frac{a\rho N_R^2}{b\rho N_R+c
N_R+1}\right),\label{app3}
\end{eqnarray}
\end{figure*}
where $W$ is a half of the spectral bandwidth, since a complete
transmission requires two time slots. $\mathcal{R}(x)$ denotes the
real part of $x$. $\textbf{h}_{R,D}$ is replaced by
$\sqrt{\rho}\hat{\textbf{h}}_{R,D}+\sqrt{1-\rho}\textbf{e}$ in
(\ref{app1}). Eq. (\ref{app2}) follows from the fact that
$\rho\|\hat{\textbf{h}}_{R,D}\|^2$ scales with the order
$\mathcal{O}(\rho N_R)$ as $N_R\rightarrow\infty$ while
$2\sqrt{\rho(1-\rho)}\mathcal{R}(\textbf{e}^H\hat{\textbf{h}}_{R,D})
+(1-\rho)\|\textbf{e}\hat{\textbf{h}}_{R,D}^H\|^2/\|\hat{\textbf{h}}_{R,D}\|^2$
scales as the order $\mathcal{O}(\sqrt{N_R})$, which can be
negligible. Eq. (\ref{app3}) holds true because of
$\lim\limits_{N_R\rightarrow\infty}\frac{\|\hat{\textbf{h}}_{R,D}\|^2}{N_R}=1$
and
$\lim\limits_{N_R\rightarrow\infty}\frac{\|\textbf{h}_{S,R}\|^2}{N_R}=1$,
namely channel hardening \cite{ChannelHardening}. Therefore, we
get the Theorem 1.

\section{Proof of Theorem 2}
According to (\ref{eqn5}), for a given $\varepsilon$, we have
\begin{eqnarray}
\varepsilon&=&P_r\left(C_{SOC}^{AF}>C_{D}^{AF}-W\log_2(1+\gamma_{E}^{AF})\right)\nonumber\\
&=&P_r\left(\gamma_E^{AF}>2^{\left(C_{D}^{AF}-C_{SOC}^{AF}\right)/W}-1\right)\nonumber\\
&=&1-F\left(2^{\left(C_{D}^{AF}-C_{SOC}^{AF}\right)/W}-1\right),\label{app4}
\end{eqnarray}
where $F(x)$ is the cumulative distribution function (cdf) of
$\gamma_E^{AF}$. In order to derive the secrecy outage capacity, the
key is to get the cdf of $\gamma_E^{AF}$. Examining (\ref{eqn11}),
due to channel hardening, we have
\begin{eqnarray}
\gamma_E^{AF}=\frac{dN_R\left|\textbf{h}_{R,E}^H\frac{\hat{\textbf{h}}_{R,D}}{\|\hat{\textbf{h}}_{R,D}\|}\right|^2}
{e\left|\textbf{h}_{R,E}^H\frac{\hat{\textbf{h}}_{R,D}}{\|\hat{\textbf{h}}_{R,D}\|}\right|^2+cN_R+1}.\label{app5}
\end{eqnarray}
Since $\hat{\textbf{h}}_{R,D}/\|\hat{\textbf{h}}_{R,D}\|$ is an
isotropic unit vector and independent of $\textbf{h}_{R,E}$,
$\left|\textbf{h}_{R,E}^H\hat{\textbf{h}}_{R,D}/\|\hat{\textbf{h}}_{R,D}\|\right|^2$
is $\chi^2$ distributed with 2 degrees of freedom. Let
$y\sim\chi_2^2$, we can derive the cdf of $\gamma_E^{AF}$ as
\begin{eqnarray}
F(x)&=&P_r\left(\frac{dN_Ry}{ey+cN_R+1}\leq x\right).\label{app6}
\end{eqnarray}
If $x<dN_R/e$, then we have
\begin{eqnarray}
F(x)&=&P_r\left(y\leq\frac{(cN_R+1)x}{dN_R-ex}\right)\nonumber\\
&=&1-\exp\left(-\frac{(cN_R+1)x}{dN_R-ex}\right).\label{app7}
\end{eqnarray}
Since $x\geq N_R/e$ is impossible when
$x=2^{\left(C_{D}^{AF}-C_{SOC}^{AF}\right)/W}-1$, we have
\begin{equation}
\varepsilon=\exp\left(-\frac{(cN_R+1)\left(2^{\left(C_{D}^{AF}-C_{SOC}^{AF}\right)/W}-1\right)}
{dN_R-e\left(2^{\left(C_{D}^{AF}-C_{SOC}^{AF}\right)/W}-1\right)}\right).\label{app8}
\end{equation}
Hence, we get the Theorem 2.

\section{Proof of Theorem 3}
In an LS-MIMO DF relaying system, the eavesdropper channel capacity can be expressed as
\begin{equation}
C_E^{DF}=W\log_2(1+\min(P_S\alpha_{S,R}N_R,P_R\alpha_{R,E}\Big|\textbf{h}_{R,E}^H\frac{\hat{\textbf{h}}_{R,D}}{\|\hat{\textbf{h}}_{R,D}\|}\Big|^2)).
\end{equation}
Then, the secrecy outage probability $\varepsilon$ with respect to a
secrecy outage capacity $C_{SOC}^{DF}$ can be computed as (\ref{app9}) at the top of the next page,
\begin{figure*}
\begin{eqnarray}
\varepsilon&=&P_r(C_{SOC}^{DF}>C_D^{DF}-C_E^{DF})\nonumber\\
&=&P_r\left(\min\left(P_S\alpha_{S,R}N_R,P_R\alpha_{R,E}\left|\textbf{h}_{R,E}^H\frac{\hat{\textbf{h}}_{R,D}}{\|\hat{\textbf{h}}_{R,D}\|}\right|^2\right)>2^{(C_D^{DF}-C_{SOC}^{DF})/W}-1\right)\nonumber\\
&=&P_r\left(P_S\alpha_{S,R}N_R\leq
P_R\alpha_{R,E}\left|\textbf{h}_{R,E}^H\frac{\hat{\textbf{h}}_{R,D}}{\|\hat{\textbf{h}}_{R,D}\|}\right|^2\right)P_r\left(P_S\alpha_{S,R}N_R>2^{(C_D^{DF}-C_{SOC}^{DF})/W}-1\right)\nonumber\\
&&+P_r\left(P_S\alpha_{S,R}N_R>
P_R\alpha_{R,E}\left|\textbf{h}_{R,E}^H\frac{\hat{\textbf{h}}_{R,D}}{\|\hat{\textbf{h}}_{R,D}\|}\right|^2\right)P_r\left(P_R\alpha_{R,E}\left|\textbf{h}_{R,E}^H\frac{\hat{\textbf{h}}_{R,D}}{\|\hat{\textbf{h}}_{R,D}\|}\right|^2>2^{(C_D^{DF}-C_{SOC}^{DF})/W}-1\right)\nonumber\\
&=&\exp\left(-\frac{P_S\alpha_{S,R}N_R}{P_R\alpha_{R,E}}\right)+\left(1-\exp\left(-\frac{P_S\alpha_{S,R}N_R}{P_R\alpha_{R,E}}\right)\right)\exp\left(-\frac{2^{(C_D^{DF}-C_{SOC}^{DF})/W}-1}{P_R\alpha_{R,E}}\right)\label{app21}\\
&\approx&\exp\left(-\frac{2^{(C_D^{DF}-C_{SOC}^{DF})/W}-1}{P_R\alpha_{R,E}}\right),\label{app9}
\end{eqnarray}
\end{figure*}
where (\ref{app21}) follows from the fact that
$\Big|\mathbf{h}_{R,E}^H\frac{\hat{\mathbf{h}}_{R,D}}{\|\hat{\mathbf{h}}_{R,D}\|}\Big|^2$
is $\chi^2$ distributed with 2 degrees of freedom, and (\ref{app9})
holds true since
$\exp\left(-\frac{P_S\alpha_{S,R}N_R}{P_R\alpha_{R,E}}\right)$
approaches zero when $N_R$ is sufficient large. Based on (\ref{app9}), it is easy to get the Theorem 3.

\section{Proof of Theorem 4}
If $P_S$ is sufficiently large, the secrecy outage capacity with
respect to a given outage probability $\varepsilon$ in an LS-MIMO AF
relaying system can be approximated as
\begin{eqnarray}
C_{SOC}^{AF}&=&W\log_2\left(1+\frac{P_SP_R\alpha_{S,R}\alpha_{R,D}\rho
N_R^2}{P_R\alpha_{R,D}\rho
N_R+P_S\alpha_{S,R}N_R+1}\right)\nonumber\\
&&-W\log_2\left(1+\frac{P_SP_R\alpha_{S,R}\alpha_{R,E}N_R\ln\varepsilon}{P_R\alpha_{R,E}\ln\varepsilon-1-P_S\alpha_{S,R}N_R}\right)\nonumber\\
&\approx&W\log_2\left(1+\frac{P_SP_R\alpha_{S,R}\alpha_{R,D}\rho
N_R^2}{P_S\alpha_{S,R}N_R}\right)\nonumber\\
&&-W\log_2\left(1+\frac{P_SP_R\alpha_{S,R}\alpha_{R,E}N_R\ln\varepsilon}{-P_S\alpha_{S,R}N_R}\right)\label{app10}\\
&\approx&W\log_2(P_R\alpha_{R,D}\rho
N_R)\nonumber\\&&-W\log_2(-P_R\alpha_{R,E}\ln\varepsilon)\label{app11}\\
&=&W\log_2\left(-\frac{\alpha_{R,D}\rho
N_R}{\alpha_{R,E}\ln\varepsilon}\right),\label{app12}
\end{eqnarray}
where (\ref{app10}) follows the fact that the terms
$``P_R\alpha_{R,D}\rho N_R+1"$ and
$``P_R\alpha_{R,E}\ln\varepsilon-1"$ in the denominators is
negligible when $P_S$ is large enough, and (\ref{app11}) holds true
since the constant ``1" is dropper in the case of large $P_S$. From
(\ref{app12}), it is known that the secrecy outage capacity is
independent of transmit power $P_S$ and $P_R$, and reduces as the
interception distance and the outage probability decreases, which
proves Theorem 4.

\section{Proof of Theorem 5}
In the case of large $P_R$, the secrecy outage capacity with respect
to a given outage probability $\varepsilon$ in an LS-MIMO AF
relaying system can be approximated as
\begin{eqnarray}
C_{SOC}^{AF}&=&W\log_2\left(1+\frac{P_SP_R\alpha_{S,R}\alpha_{R,D}\rho
N_R^2}{P_R\alpha_{R,D}\rho
N_R+P_S\alpha_{S,R}N_R+1}\right)\nonumber\\
&&-W\log_2\left(1+\frac{P_SP_R\alpha_{S,R}\alpha_{R,E}N_R\ln\varepsilon}{P_R\alpha_{R,E}\ln\varepsilon-1-P_S\alpha_{S,R}N_R}\right)\nonumber\\
&\approx&W\log_2\left(1+\frac{P_SP_R\alpha_{S,R}\alpha_{R,D}\rho
N_R^2}{P_R\alpha_{R,D}\rho
N_R}\right)\nonumber\\
&&-W\log_2\left(1+\frac{P_SP_R\alpha_{S,R}\alpha_{R,E}N_R\ln\varepsilon}{P_R\alpha_{R,E}\ln\varepsilon}\right)\label{app13}\\
&\approx&W\log_2(P_S\alpha_{S,R}N_R)-W\log_2(P_S\alpha_{S,R}N_R)\label{app14}\\
&=&0,\label{app15}
\end{eqnarray}
where (\ref{app13}) follows the fact that the terms
$``P_S\alpha_{S,R}N_R+1"$ and $``P_R\alpha_{R,E}\ln\varepsilon-1"$
in the denominators is negligible when $P_R$ is large enough, and
(\ref{app14}) holds true since the constant ``1" is dropper in the
case of large $P_R$. From (\ref{app15}), it is found that the
secrecy outage capacity is always equal to zero with arbitrary
transmit powers $P_S$ and $P_R$, interception distance and outage
probability, which proves Theorem 5.

\section{Proof of Theorem 6}
If $P_S$ is sufficiently large, the secrecy outage capacity with
respect to a given outage probability $\varepsilon$ in an LS-MIMO DF
relaying system can be computed as
\begin{eqnarray}
C_{SOC}^{DF}&=&W\log_2\left(1+\min(P_S\alpha_{S,R}N_R,P_R\alpha_{R,D}\rho
N_R)\right)\nonumber\\&&-W\log_2\left(1-P_R\alpha_{R,E}\ln\varepsilon\right)\nonumber\\
&=&W\log_2\left(1+P_R\alpha_{R,D}\rho
N_R\right)\nonumber\\&&-W\log_2\left(1-P_R\alpha_{R,E}\ln\varepsilon\right)\label{app16}\\
&=&W\log_2\left(\frac{1+P_R\alpha_{R,D}\rho
N_R}{1-P_R\alpha_{R,E}\ln\varepsilon}\right),\label{app17}
\end{eqnarray}
where (\ref{app16}) follows the fact of
$\min(P_S\alpha_{S,R}N_R,P_R\alpha_{R,D}\rho
N_R)=P_R\alpha_{R,D}\rho N_R$ at a large $P_S$. From (\ref{app17}),
it is found that the secrecy outage capacity is independent of
$P_S$, and declines as the interception distance the outage
probability decrease. Furthermore, if $P_R$ is relatively large, the
secrecy outage capacity can be approximated as
\begin{eqnarray}
C_{SOC}^{DF}&=&W\log_2\left(-\frac{\alpha_{R,D}\rho
N_R}{\alpha_{R,E}\ln\varepsilon}\right),\label{app18}
\end{eqnarray}
where (\ref{app18}) holds true since the constant ``1" in both the
denominator and the numerator is negligible when $P_R$ is high.
Interestingly, it is found that DF has the same secrecy outage
capacity as AF, which proves the claims in Theorem 6.

\section{Proof of Theorem 7}
In the case of large $P_R$, the secrecy outage capacity with respect
to a given outage probability $\varepsilon$ in an LS-MIMO DF
relaying system can be approximated as
\begin{eqnarray}
C_{SOC}^{DF}&=&W\log_2\left(1+\min(P_S\alpha_{S,R}N_R,P_R\alpha_{R,D}\rho
N_R)\right)\nonumber\\&&-W\log_2\left(1+P_R\alpha_{R,E}\ln\varepsilon\right)\nonumber\\
&=&W\log_2\left(1+P_S\alpha_{S,R}N_R\right)\nonumber\\&&-W\log_2\left(1-P_R\alpha_{R,E}\ln\varepsilon\right)\label{app19}\\
&=&0,\label{app20}
\end{eqnarray}
where (\ref{app19}) follows the fact of
$\min(P_S\alpha_{S,R}N_R,P_R\alpha_{R,D}\rho
N_R)=P_S\alpha_{S,R}N_R$ at a large $P_R$, and (\ref{app20}) holds
true because of $C_{SOC}^{DF}\geq0$. From (\ref{app20}), it is found
that the secrecy outage capacity is always equal to zero with
arbitrary transmit powers $P_S$ and $P_R$, interception distance and
outage probability. In addition, DF has the same secrecy outage
capacity as AF, which proves Theorem 7.

\end{appendices}

\end{document}